# RF-Flashlight Testbed for Verification of Real-Time Geofencing of EESS Radiometers and Millimeter-Wave Ground-to-Satellite Propagation Models[1]

Elliot Eichen, Arvind Aradhya, and Ljiljana Simić[2]

*Abstract*— A simple "RF-flashlight" (or ground-to-satellite) interference testbed is proposed to experimentally verify (i) real-time geofencing (RTG) for protecting passive Earth Exploration Satellite Services (EESS) radiometer measurements from 5G/6G mm-wave transmissions, and (ii) ground-to-satellite propagation models used in the interference modeling of this spectrum coexistence scenario. RTG is a stronger EESS protection mechanism than the current methodology recommended by the ITU based on a worst-case interference threshold while simultaneously enabling dynamic spectrum sharing and coexistence with 5G/6G wireless networks. Similarly, verifying more sophisticated RF propagation models that include ground topology, buildings, and non-line-of-sight paths will provide better estimates of interference than the current ITU line-of-sight model and, thus, a more reliable basis for establishing a consensus among the spectrum stakeholders.

## I. INTRODUCTION

The ever-growing demand for wireless services has created competition for RF spectrum between wireless communications networks and legacy applications such as remote sensing, GPS, radio astronomy, and radar systems. The potential interference between 5G/6G transmissions and passive Earth Exploration Satellite Services (EESS) measurements has been particularly problematic when allocating mm-wave spectrum because of the extreme sensitivity of EESS radiometers and the critical importance of their measurements to weather forecasting and climate modeling balanced against the potential for fiber-equivalent bandwidths delivered by wireless networks [1,2].

The issue of interference between these two uses of RF spectrum is also clouded by the methodology used to estimate this interference; the communications community claims that the model adopted by the International Telecommunications Union (ITU) used to calculate interference is too restrictive and that it overestimates the amount of interference. In contrast, the remote sensing and weather communities claim ITU adopted requirements based on this model (e.g., ITU Recommendations from the 2019 World Radio Congress (WRC) for out-of-band emissions into the 23.8GHz observation band [3]) underestimate the interference from 5G networks. When layered on top of a diplomat-led political process (the WRC) for setting and adopting requirements, these technical disagreements reinforce an environment of mistrust where both sides believe the ITU recommendations are incorrect and harmful to their interests.

This paper proposes a simple "RF-flashlight" testbed (Figs. 1 and 2) to observe changes in EESS radiometer measurements when ground transmissions are turned ON from OFF. This testbed would be used to experimentally verify the feasibility of real-time geofencing (RGT) [4], a dynamic coexistence mechanism that enables 5G/6G networks to share and more fully exploit mm-wave spectrum while simultaneously protecting EESS weather-prediction radiometers. RGT works by pausing network transmissions (and potentially moving traffic to other bands) in the "dark-spaces" when EESS measured pixels subtend transmission antennas, while allowing transmissions in the "white-spaces". It is thus a stronger yet more flexible protection mechanism than the fixed threshold of maximum allowed worst-case interference adopted by the ITU.

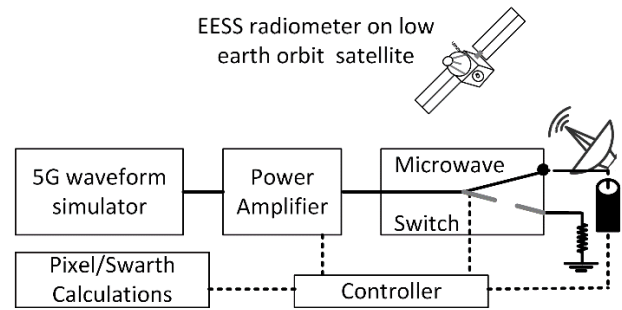

Fig. 1: RF-flashlight testbed configuration including a fast microwave switch.

Furthermore, the RF-flashlight testbed can be used to directly measure the interference generated at the EEES radiometers by terrestrial communication networks, thereby verifying more realistic radio frequency propagation calculations that go beyond the ITU's simplistic Line-of-Sight (LOS) model by incorporating ground topology, buildings, and the resulting reflected non-LOS (NLOS) propagation paths, as well as the directional nature of beamformed 5G/6G transmissions. We emphasize that verifying such propagation models for the ground-satellite interference channel is crucial to enable a better consensus among the stakeholders in this spectrum coexistence. More reliable interference estimates for planned 5G/6G wireless deployments, in turn, benefit the development, testing, and deployment of RGT, as well as other candidate interference mitigation mechanisms.

[1] This material is based in part upon work supported by the National Science Foundation under Grant No. 2232368
[2] Elliot Eichen (elliot.eichen@colorado.edu) and Arvind Aradhya (arvind.aradhya@colorado.edu), Computer Science Dept, University of Colorado Boulder, Boulder CO USA. Ljiljana Simić (lsi@inets.rwth-aachen.de), Institute for Networked Systems, RWTH Aachen University, Aachen, Germany.

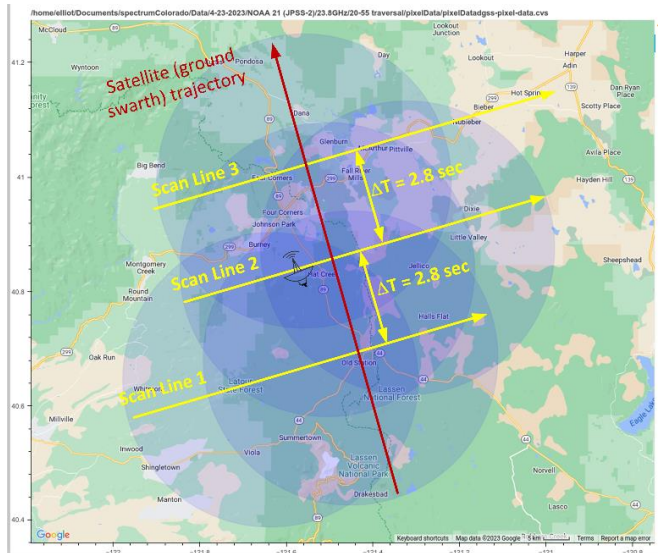

Fig. 2: Using radiometers with overlapping pixels in adjacent scan lines enables verification of real-time geofencing and RF propagation models without correcting for satellite position, atmospheric environment, or different sensors.

Whereas the proposed testbed could be used to test a variety of propagation models and interference mitigation techniques, this paper focuses on:

i.) Using actual (deployed) 5G network infrastructure (type, location, pointing angles, etc.) and transmitter characteristics (power levels, antenna radiation patterns, electrical spectrums, etc.) rather than the ITU's Monte-Carlo simulation of network infrastructure and theoretical transmitter characteristics. This includes network components included in the ITU interference models (base stations (gNBs) and user equipment (UEs)) as well as components not considered by the ITU (such as Integrated Access and Backhaul (IAB) repeaters[6] and high power (3GPP Class V) UEs[3] [7]).

ii.) Employing RTG to protect EESS assets by pausing or limiting traffic based on class-of-service and/or moving traffic to alternative bands. This contrasts with the ITU's methodology, which requires all network equipment to meet a predetermined worst-case (0.01% of all measurement pixels) interference scenario based on assumptions of network infrastructure and architecture [8].

iii.) Testing more accurate mm-wave RF propagation algorithms that use the actual (deployed) network infrastructure, building and topographical data, and ray tracing to include reflections and other NLOS transmission paths [5,9].

The rest of this paper is organized as follows. Section II reviews the ITU interference calculation methodology and interference management requirements. Section III presents RTG as an alternative dynamic interference management technique. Section IV discusses more sophisticated propagation models for the ground-to-satellite mm-wave channel to allow for more reliable and realistic interference estimates. Section V presents our proposal for an RF-flashlight testbed to experimentally verify RTG and the propagation models used in interference calculations. Section VI concludes the paper.

## II. ITU INTERFERENCE MODEL & MITIGATION PROCESS

Reviewing the ITU's model and process for managing interference is helpful in contrasting RTG and more sophisticated propagation models and interference calculation methodologies (*cf*. Sections III and IV). To mitigate 5G/6G mm-wave network interference with EESS radiometer measurements, the ITU requires that all individual network and user equipment transmitters must limit their interference levels so that the aggregate interference from all transmitters that subtend any given radiometer measurement pixel is less than -200 dBm/MHz for 99.99% of all such measurements within a $2 \times 10^6$ km$^2$ area (Fig. 3). The ITU methodology for calculating this interference [10] defines the network elements as 3GPP Class I 5G gNBs, the UEs as 3GPP Class IV mobile handsets, and the 5G/6G antenna radiation patterns based on MIMO theoretical models. The ITU further defines the 5G/6G infrastructure as a Monte Carlo simulation of gNB/UE density and parameters within three different scenarios (urban, suburban, and rural) and provides a list of radiometer sensors [8] with relevant parameters (antenna gain, NE$\Delta$T (noise equivalent delta temperature) of the radiometer, etc.) as input to the Monte Carlo simulation. The ITU's propagation model only assumes LOS transmission and does not account for buildings, topography, reflections, or other sources of multipath propagation. It is important to note that the ITU methodology for estimating interference is based on an entirely theoretical description of the communication system's architecture, locations, and devices.

Given the above methodology for calculating interference, it is then up to member countries to run the Monte Carlo based simulations to estimate interference limits for 5G/6G deployments. Even though the inputs and the methodology are relatively well defined, the fact that member countries "bake" their own simulations based on their own "ingredients" and the ITU "recipe" nonetheless often leads to diverging estimates of interference [11]. This, in turn, further complicates negotiations of the "correct" per-network-element (i.e., gNBs and UEs) interference limits that ought to be mandated to protect EESS radiometer measurements. Moreover, while the results of these different estimates are eventually available from the ITU's website, details of the simulation and the computer code used to calculate interference are the property of the individual nations. They are typically not made available for peer review [12].

---
[3] designed for fixed wireless access (FWA) UEs that enable wireless broadband distribution with fiber-like speeds.

From a timing perspective, the ITU process is long and cumbersome, typically requiring two four-year ITU WRC cycles between service and spectrum proposals to ratify interference recommendations. Moreover, equipment and underlying chip vendors need one to two additional years before products become available. Even an accelerated process (one four-year ITU WRC cycle plus one year for hardware and chip development) would be extraordinarily long compared with the rapid pace of technological change in wireless communications. This asymmetry between the ITU process and technology evolution inevitably leads to cases where the ITU interference recommendations are out-of-date even before ratification (as happened in WRC-19[13]). Moreover, it also results in commercial uncertainty that delays investment into hardware and chip development and, thus, service deployments.

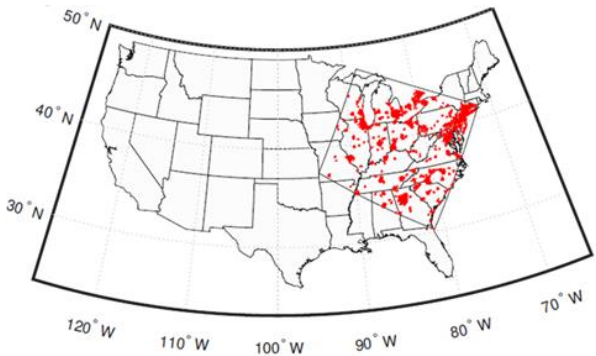

Fig. 3. The ITU model requires the aggregate interference from transmitters that subtend a measurement pixel to contribute less than -200 dBm/MHz for 99.99% of all pixels within a $2 \times 10^6$ km$^2$ boundary. From [10].

### III. REAL-TIME GEOFENCING

In contrast to the ITU's hardware-based static methodology for preventing interference, RTG is a software-based dynamic interference mitigation methodology that enables spectrum sharing by allowing communication networks to operate in the "white-spaces" between ESS radiometer measurements while protecting EESS measurements in the "dark-spaces." For the specific case of EESS measurements at 23.8 GHz experiencing out-of-band interference from 5G FR2 networks at 24 GHz, the ratio of white-space times to dark-space times is greater than 99:1 when aggregated across all 21 satellites/radiometer pairs making measurements at 23.8 GHz [4][4]. Simply pausing transmissions during the dark space thus results in a network availability of ~99% [4]. Adding the ability to move 5G traffic between bands during a dark space (which leverages functionality included in 2022 3GGP Release 16) could in principle, improve the network availability to greater than 99.99%, enabling those 5G networks to provide better coverage and wider bandwidths by operating at higher (3PGG Class V) transmission powers in the dark-spaces and exploiting mm-wave spectrum.

While RTG has been discussed previously [4,14], for the purposes of this paper, it is sufficient to consider that RTG and better propagation models have significant advantages to the existing ITU models for both the protection of EESS radiometer measurements and for the coverage, performance, and spectrum sharing with 5G/6G mm-wave networks (Table 1 – p5). *Adopting RTG or other spectrum protection and sharing mechanisms and improved propagation and interference models would benefit from direct experimental verification to build stakeholder confidence and consensus.*

### IV. IMPROVED RF PROPAGATION MODELS

Recent works [5,9] have considered more sophisticated propagation models for the ground-to-satellite interference channel in order to better estimate the level and spatial structure of interference from emerging 5G/6G cellular deployments to passive sensing EESS systems. These propagation models incorporate both LOS and reflected NLOS paths by considering the geometry of site-specific deployment scenarios, incorporating reflections off buildings [5] and the ground [5,9]. Moreover, the interference analyses in [3] and [9] both consider realistic ground network node locations and densities as well as realistic beamforming configurations (in terms of antenna radiation pattern and pointing angles) for the ground gNB/UE transmitter as well as the scanning satellite.

The authors in [5] presented a detailed coexistence study of 5G mm-wave networks and weather satellites sensing in the 23.8 GHz band. They adopted a 3D geometric model for the ground-satellite channel, consisting of LOS and NLOS propagation paths from strong (up to six bounces) reflections off the ground or surrounding buildings, and obtained site-specific propagation data for their study area in New York City using ray tracing simulations with real 3D building data. They also assumed realistic directional antenna patterns and beam pointing angles, thus taking into account potential interference originating from both the main lobe and sidelobes of the ground transmitter. The results of their interference analysis showed that the highest interference levels at the satellite from uplink and downlink transmissions were due to multiple strong NLOS reflections off building walls and the ground, respectively. This underlies the importance of incorporating NLOS propagation to properly estimate the ground-to-satellite interference and the inadequacy of LOS-only modeling approaches as adopted by the ITU.

The authors in [9] presented a general coexistence analysis of 6G networks and passive sensing satellites in the sub-THz bands,

---

[4] The dark spaces include significant buffering to account for spatial frequency (Nyquist) constraints and also full scan line isolation for all radiometers (except ATMS) as the pointing angle of the radiometer (i.e. the rotation angle of the RF reflector) is not synchronized to the satellite position for all radiometers except ATMS. In these cases, a "dark space" is really a "dark scan line."

with case studies at 164, 178, and 240 GHz using large-scale realistic 6G network topologies in Boston based on real street layout and 3D building data. The authors explicitly compared the estimated interference considering only the LOS component (as per the ITU model) against their two-ray model, i.e. both LOS and the ground reflection NLOS ray. They showed that the ground reflection NLOS path contributes significant interference due to the geometry of the ground-satellite interference channel and thus cannot be neglected. They also noted that the attenuation outside the main lobe of the 6G directional antenna arrays may be insufficient to protect passive users. This emphasizes that NLOS interference paths via sidelobes cannot be neglected in these coexistence scenarios, consistent with the findings from [35].

Although both of these works rely on more complex propagation models that strive to incorporate pertinent details of the site-specific propagation environment, they nonetheless still rely on theoretical models and simulated propagation data (using ray-tracing-like frameworks). Namely, the propagation models used in [5,9] have still not been validated against real measurement data[5] from the ground-to-satellite interference channel. It is thus not yet established to what extent these more detailed models may, in fact, over- or under-estimate interference (e.g., by poor tuning of parameters such as reflection losses, insufficiently accurate modeling of the urban/terrestrial environment, or by neglecting the impact of secondary propagation mechanisms such as diffraction and scattering). Similarly, it has not yet been determined what degree of increased modeling effort is justified by the potential corresponding gain in the interference estimation accuracy (i.e., in terms of sufficient terrain/building/antenna pattern modeling detail, *cf.* analysis for ground mm-wave networks in [16]).

It would thus be enormously beneficial to explicitly verify the accuracy of more sophisticated propagation models as adopted in [5,9] via the gold-standard method of comparing them against real RF measurement data as a "ground-truth" baseline. Controlled interference measurements using the proposed RF-flashlight testbed would enable precisely such an experimental verification of ground-to-satellite propagation models to be systematically performed and would be essential for establishing an evidence-based consensus on the reliability of the resulting interference estimates.

V. PROPOSED RF-FLASHLIGHT TESTBED: EXPERIMENTAL CONFIGURATION

The proposed RF-flashlight experimental configuration – in its simplest form – is shown in Fig. 1. A single antenna with a narrow beamwidth (e.g. a standard-gain directional horn) is mounted on a dual-axis (azimuth and elevation) rotator. A 5G waveform generator centered at frequency $f_c$ with bandwidth $\Delta f_c$ is coupled to a power amplifier and then through a fast microwave switch to the transmission antenna. To test an RTG system's ability to predict the time and location of dark spaces, the antenna can be pointed directly at a satellite radiometer in orbit such that the gains of both the RF-Flashlight's ground transmitter and the radiometer's antennas are maximized. For testing RF propagation models (including the impact of dominant LOS or NLOS interference paths), the RF-flashlight antenna can be pointed in various directions with and without various terrain or building impediments. Since the objective is to measure the anthropogenic radiation above the natural atmospheric emission, the RF-flashlight transmitter should be located in a remote area that is known to be free from other sources of radiation at $f_c \pm n\Delta f_c$ where $n$ provides reasonable buffering in the neighborhood of $f_c$.

To obtain cooperation and permission to operate this testbed, it is critical to ensure that ***any use of the RF flashlight cannot harm the proposed test radiometer*** or interfere with any other legacy spectrum user. Moreover, it is also imperative to ensure that using the RF-Flashlight has no impact on any existing numerical weather prediction *(NWP) models*. These criteria can be met by:

i.) working with the satellite operator to review the experimental configuration and audit its radiated power to ensure that the radiometer cannot be harmed;

ii.) reviewing the test location and frequencies with the NTIA and FCC; and iii.) developing the software to exclude potentially bad (corrupted) measurement pixels resulting from the operation of the RF-flashlight testbed from input to NWP systems. (This would presumably attach to the existing process of filtering such data).

The experimental methodology is to compare two measurements – one with the transmitter on and one with the transmitter off – made sufficiently close to each other in time such that natural atmospheric radiation emitted at $f_c \pm \frac{1}{2}\Delta f_c$ and the propagation characteristics between the transmitter and radiometer have not changed. One simple way to achieve this is to use a radiometer where subsequent scan lines overlap such that the transmitting antenna is subtended by pixels in multiple scan lines (Fig. 2) separated by the cycle time of the radiometer (typically a few seconds), and where the phase of the scanning mechanism at time $t + \tau$ can be predicted from the phase at time $t$ (i.e., the radiometer scanning mechanism is run in closed loop with respect to the satellite's position). In this case, an RF Flashlight measurement (received power with Flashlight ON) can be compared with overlapping pixels in a subsequent scan line (received power with Flashlight OFF). In this case, the ON time or the flashlight would be on the order of a hundred milliseconds at most for ATMS.

---

[5] The ray tracing propagation simulator used in [3] has been validated against real-world mm-wave outdoor measurements in [16] but only for ground-to-ground mm-wave links.

A second way to perform these experiments is to use two "identical" radiometers flying on satellites in similar orbits separated by a short time. This second method is less ideal, particularly for testing RF propagation models that require quantitative measurements of the received power, since this method requires corrections due to the change in location of the satellite with respect to the signal transmitter and any potential corrections necessitated by using different sensors. If the radiometer, in this case, did not allow phase recovery of the scanning position, geofencing would have to take place on a scan line basis, which means an ON time equal to a few seconds.

To get a sense of the power required by the RF-flashlight transmitter, assume that the EESS radiometer has overlapping scan lines, so that no correction is required for change in satellite location or atmospheric conditions. If:

- $P_{ON}$ = the transmit power from the RF-flashlight,

- $P_{interference}$ = the total RF interference power from the RF-flashlight at the radiometer sensor
  $= L\, P_{ON}\, G_{transmitter}(\phi,\theta)\, G_{radiometer}(\phi,\theta)$,

- $P_{noise}$ = the noise equivalent temperature of the sensor
  $= k_b N_{temp} \Delta f$ ($k_b$ is Boltzmann's constant), and

- $P_{H2O}$ = the RF power at the sensor atmospheric water vapor.

where $L = L_{path-loss}\, L_{atmosphere}\, L_{polarization}$ is the total loss between radiometer and transmitter due propagation path loss (including both LOS and NLOS propagation paths), atmospheric attenuation, and sensor polarization.

The ratio of RF power detected by the radiometer when the signal transmitter is *on* ($EESS_{on}$) and *off* ($EESS_{off}$) is then given by

$$\frac{EESS_{on}}{EESS_{off}} = 1 + \frac{L\, P_{ON}}{P_{noise} + P_{H2O}}$$

Using typical parameters $N_{temp}, \Delta f, L_{atmosphere}, L_{free-space}$, and $L_{polarization}$ for an example radiometer and traversal (Table 1), taking $P_{H2O} \sim 100 P_{noise}$, and somewhat arbitrarily requiring that the measured signal $EESS_{on}$ has to be 10 times greater than $EESS_{off}$, then $P_{ON} \sim 10$ Watts and the received power from the RF-Flashlight at the radiometer is $\sim -110$ dBm. These are reasonable values, and *well below RF power levels that could damage EESS radiometers in Low Earth Orbit*.

VI. SUMMARY

The ever-growing demand for wireless services has created competition for RF spectrum between wireless networks and legacy applications. The potential interference between 5G/6G transmissions and EESS radiometer measurements has been particularly problematic when allocating mm-wave spectrum because of the extreme sensitivity of EESS radiometers and the critical importance of their measurements to weather forecasting and climate modeling balanced against the potential for fiber-equivalent bandwidths delivered by wireless networks.

| Parameter | scan position | |
|---|---|---|
| | Nadir | Edge |
| sat name | NOAA-21 (JPSS-2) | |
| NORAD ID | 54234 | |
| date and traversal time | 4/23/2023 | |
| sat lat | 40.774 | 40.828 |
| sat long | -120.988 | -121.006 |
| sat altitude [km] | 832.1 | 832.1 |
| pixel lat | 40.646 | 42.701 |
| pixel long | -121.637 | -105.927 |
| pixel altitude [m] | 20 | 20 |
| radiometer name | ATMS | ATMS |
| ITU sensor # [1] | F5 | F5 |
| center frequency [GHz] | 23.8 | 23.8 |
| bandwidth [GHz] | 0.2 | 0.2 |
| elevation angle [degrees] | 85.6 | 25.8 |
| polarization loss [dB] | -3 | -3 |
| free space loss [dB] | -185 | -184 |
| atmosphere [dB] | -6.1 | -10.9 |
| radiometer antenna max gain [dBi] [1] | 30 | 30 |
| transmitter antenna max gain [dBi] | 15 | 15 |
| total loss [dB] | -149 | -153 |

Table 2: Parameters used to estimate RF-flashlight power for transmitter and pixels across the Hat Creek Radio Observatory [13].

Real Time Geofencing (RTG) has previously been described as a practical method to increase the protection of EESS radiometers while simultaneously enabling spectrum sharing and improving the performance of terrestrial communication networks. Similarly, more sophisticated RF propagation models that include Non-Line-of-Sight (NLOS) multipaths have been developed that enable more accurate estimates of interference again enabling better protection of EESS assets while facilitating spectrum sharing. These models show that interference from NLOS paths can be substantial, but they are not currently considered when setting international interference limits.

A relatively simple RF-Flashlight (or ground-to-satellite) testbed is proposed to experimentally validate RTG and to measure interference to an EESS radiometer in low earth orbit from terrestrial communication systems. Using a single ground base transmitter and an operational radiometer currently in orbit, the testbed could make the required measurements without impacting the ability of the radiometer to provide data for numerical weather prediction models and without harming the instrument. Direct tests of the RF propagation models used to set international interference limits (which have not been done previously) would improve confidence in the ITU recommendations and alleviate some of the tension between the wireless and the weather/climate communities. Similarly, verification of RTG would support better protection for EESS assets than existing mechanisms while simultaneously enabling spectrum sharing and improved operation of 5G/6G networks.

|  | Real-Time Geofencing | ITU Methodology |
|---|---|---|
| Communications Network Performance | Optimize network performance and coverage by topology, buidling environment, and actual deployed infrastructure. "White-space" to "dark-space" ratio ~ 99:1. Increases in service area and network performance encorage spectrum sharing. Impact to network availablity in "dark-spaces" can be mitigated by using traffic migration capabilities (3GPP rel 16). | Nework and User Equipment designed for worst-case conditions based on an assumed network architecture, assumed sounder sensitivity, Monte Carlo simulations of infrastructure, and line-of-sight propegation model. Negative impact on network coverage and performance. Designing for "worst-case" scneario not synergistic with spectrum sharing. Inteference hardware based bandwidth filtering only supports adjacent bands |
| Protection of EESS Assets | EESS assets can be protected to desired interference levels by increasing dark-space coverage area. EESS assets can be adjusted based on radiometer sensitivity. | Continued debate on the short comings of the ITU model leads to inadaquite protection of EESS sounders and overly constrained 5G/6G networks |
| Integration | Third party Spectrum Access System requires integration with wireless carrier networks (Radio Access Networks) and infrastructure data. | Compliance with emissions standards embedded in network and user equipment hardware. |
| Extensibility | Single software system supports spectum sharing and EESS Asset protection across multiple mm & sub-mm wave bands. Ability to re-use existing Spectrum Access System architecture and software developed for Citizens Band Radio Service (CBRS). | Hardware solution that requires inteference and infrastructure modeling + network/user equipment design. Development and manufacturing required for each service and spectral band allocation. |
| Flexibility | Software solution quickly accomdates changes and advances in 5G/6G architectures, network elements, and user equipment. Software solution quickly accomdates improvements in EESS sounder sensitvity, methodology (pixel size etc.), or number and capabilities of satellites. | Once hardware has been deployed, it typically can not be modified. For example, improvements in sounder sensitivity, or changes in network architecture (for example, the use of IAB repeaters) cannot be accomodated without deployment of new hardware. |
| Regulatory Compliance | Compliance policing/teting based on system audits and simple field spot testing. | In-situ EIRP measurements and/or other field measurements variable and difficult. Reliance on vendor testing of pilot device in far field chambers. |

Table 1 – Comparison of RTG and ITU methodologies for EESS protection and spectrum sharing.